\documentclass[preprint,12pt]{elsarticle}
\usepackage{amssymb}
\usepackage{epsfig}
\usepackage{amsmath}
\usepackage{amssymb}
\usepackage{graphicx}
\usepackage{caption}
\usepackage{subfigure}

\journal{Physica A}

\begin{document}

\begin{frontmatter}

\title{The Effect of Social Media on Shaping Individuals Opinion
  Formation}

\author{{ Semra G\"und\"u\c{c}}}

\address{Department of Computer Engineering \\
        Faculty of Engineering, Ankara University\\
        06345 G\"olbas{\i} Ankara, Turkey\\
        {email:{gunduc@ankara.edu.tr}}}

\begin{abstract}
In this paper, the influence of the social media on the opinion
formation process is modeled during an election campaign. In the
proposed model, peer-to-peer interactions and targeted online
propaganda messages are assumed to be the driving forces of the
opinion formation dynamics. The conviction power of the targeted
messages is based on the collected and processed private information.
In this work, the model is based on an artificial society, initially
evenly divided between two parties. The bounded confidence model
governs peer-to-peer interactions with a value of confidence parameter
which leads to consensus. The targeted messages which was modeled as
an external interacting source of information convert some weakly
committed individuals to break this evenness. Both parties use the
same methods for propaganda. It is shown that a very small external
influence break the evenness of the opinion distribution which play
significant role in the election results.  Obtained opinion
fluctuation time series have close resemblance with the actual
election poll results.
\end{abstract}

\end{frontmatter}

\section{Introduction}


During the last decade internet and particularly online news services
and social media networks have been the dominant information sharing
channels.  In the social media large groups of individuals, sharing
similar interests form networks~\cite{Bessi:2016} which
create mutual trust among the members of the network. Information
coming from a member of the network is accepted and propagated by the
members of the group without much criticism
\cite{Mocanu:2015,Askitas:2017,Shao:2018}. Such a free environment
make the users vulnerable since as well as expressing their opinions
they also reveal some personal data. Third parties may collect such
personal information, process it and use for their
purposes. Advertising agencies and political parties are willing use
the personal information to convince or to convert individuals. Hence
the freedom of expressing opinion and spreading information may be
misused to propagate rumor, gossip and misleading or false
information~\cite{Bakshy:2015,DelVicario:2017,Schmidt:2017}. The
question, whether such interventions  effect
the public opinion or not~\cite{Bessi:2016c} is still an open debate. The effect of an
information system such as TV, newspaper, blogs, on the evolution of
public opinion is studied in \cite{calderelli}. Also, the analysis of
the frequency of interaction is considered in \cite{frequency}. It is
also shown that, to use/support some concepts which are valuable or
very sensitive for the society, such as religion, nationality,
cultural issues, collective beliefs can also make a profit for
politicians \cite{Semra} and even the ideas adopted by the minority of
the society can be supported by the majority, after such a process.The
effect of the social network utilizing degree-dependent fitness and
attributes when there are competing opinion diffusion is introduced
in~\cite{competing}.  The role of the social media on the social
polarization of the society is studied through both
data~\cite{Bessi:2016,Bessi:2016b} and model
studies~\cite{DelVicario:2017,Askitas:2017}. Although the social
studies still do not have clear evidence on the influence of such misleading
information flow on the social preferences, it is shown that false
news propagation is faster and broader than the spread of true news
due to the attractiveness of the false news~\cite{Vosoughi:2018}.

$\;$

Recently, two very important social events, namely Brexit –the British
referendum to leave the European Union– and 2016 US presidential
election campaigns are the striking examples of such social
phenomena. During the campaign the fake news are not only fabricated
but more than that the issues are carefully selected by using personal
information of the social media users~\cite{CA,Bovet:2019}. The collected personal
information is used to convince and convert individuals. It is shown
that in the US Presidential election campaigns some nodes and bots,
i.e. automated accounts, in the social networks spread fabricated,
fake, biased information, distort actual news, disseminate deceptive
information~\cite{Shao:2018,Bessi:2016c,Vosoughi:2018,Ferrara:2016}.
The severity of the outside interruption can be seen by comparing the
numbers: Only on 20 election stories, the number of Facebook
engagements are 8.7m fake news versus 7.3m mainstream news starting
from the beginning of August to the election
day~\cite{Statista:FakeNews}. During a recent survey, nearly $85$
percent of respondents stated that they believed fake news is a
serious social problem~\cite{Statista:FakeNews2}.

$\;$ 

The aim of this work is to build a model to study the effects of
varying sources of fake and biased news on the opinion formation
during an election campaign.  The model society consist of $N$
individuals with multi-component opinion on the election issues. On
the decision-making process information, originating from different
sources, play a vital role to build up opinion. For the simplicity
election issues are limited in the model. The individuals exchange
opinion on only three different issues (such as the economy, health
services, security).  Hence, each individual is identified by three
real opinion values, but when it comes to making a decision on the
vote, the choice is a result of combined opinion formed on all three
issues. In this sense, the opinion structure of the model resembles
Axelrod mode \cite{Axelrod}. The individuals exchange opinion
according to the bounded confidence model (BCM)\cite{Deffuant}.
Each individual is also subject to information flow through public services, social media,  and online news sources.  Some of this information may be fake, fabricated and even targets a particular individual. Targeted news specifically
designed by considering the individual preferences~\cite{CA} which is more effective on the non-committed individuals. To mimic the effects of the public services, social media, targeted, false or biased information spread a new interaction is introduced.

situations.

$\;$

The model is tested on four different cases:
\begin{enumerate}
\item Individual interact with others through peer-to-peer interactions no external sources affects the dynamics.
\item One of the parties send messages to convert less convinced individuals.
\item Both of the parties send mes-ages to convert less convinced individuals.
 \item Both parties send messages but one sends more convincing (fabricated)  messages.  
  \end{enumerate}
In all four cases individuals exchange opinion through social media
channels.  The acceptability of the messages of online news sources
are controlled by using different probability values which are
discussed in detail in the results section.
  
The rest of this paper is organized as follows. Section 2 provides a
background for the proposed model with bounded confidence opinion
dynamics. Section 3 is devoted for presentation of the simulation
results. Finally Section 4 concludes the paper.

\section{The Model\label{Model}}

 The proposed model is based on an artificial society with
$N$ individuals. The communication network is a fully connected
network with nodes, $i=1,\dots N$. This topology allows every
individual to interact with every other mutually. Even though it looks
too simple; it eliminates the artifacts of more complicated network
topology which is essential for our discussion. The position of each
member of the society is labeled by Latin alphabets $i,j\dots$. Each
individual carries a three opinion component which is labeled by Greek
alphabet, $\alpha=1,2,3$.

Eq.\ref{opinion} defines the opinion of an individual who has three
opinion components in the matrix form.

\begin{equation}
\label{opinion}
O = \left(
\begin{array}{ccc}
o_{11} & o_{12} & o_{13}\\
o_{21} & o_{22} & o_{23}\\
\vdots&\vdots&\vdots\\
o_{N1} & o_{N2} & o_{N3}\\
\end{array}
\right)
\end{equation}

Here $N$ is the number of individuals, $o_{i \alpha}$, ($o_{i\alpha}
\in  \mathbb{R}`w$,) are the opinion values of $i^{\rm th}$ individual on
the $\alpha^{\rm th}$ issue.  All three opinion components are
assigned randomly to each individual. Each opinion component has a
Gaussian distribution with mean value $<O>=\pm 1$, indicates two
opposing views.  In order to have some interaction between different
views (different opinion individuals), variance of the Gaussian is
used as the control parameter of the overlapping region.

$\;$

At every interaction a randomly chosen pair of individuals
exchange opinion on a randomly chosen issue. In each interaction
individuals discuss on any one of the three issues in concern.
Opinion exchange is realized according to BCM \cite{Deffuant} given in
Eq.\ref{BCM}.

\begin{eqnarray}
\label{BCM}
if\; |o_{i\alpha}(t) &-& o_{j\alpha}(t)| \le \Delta \\
o_{i\alpha}(t) &=& \omega o_{i \alpha}(t-1) + ( 1- \omega ) o_{j \alpha}(t-1)\nonumber \\
o_{j\alpha}(t) &=& \omega o_{j\alpha}(t-1) + ( 1- \omega ) o_{i \alpha}(t-1) \nonumber
\end{eqnarray}
Here $\Delta$ is the tolerance threshold, $\omega$ is the opinion
exchange factor, and $t$ indicates discrete time steps.

Eq. \ref{proposedmodel} defines the interaction of external influences
with the individuals.

\begin{equation}
  {\rm \; if}\;\; o_{i \alpha} \le \eta_{i, \alpha}\;\; {\rm then} \;\; o_{i \alpha} = \left \{ \begin{array}{ll} o_{i \alpha} + M_{i \alpha} & {\rm \;\; if}\;\;  P_{i \alpha} > r  \\ & \\ o_{i \alpha} & {\rm otherwise}
    \end{array}
  \right .
  \label{proposedmodel}
\end{equation}

Here, $P_{i \alpha}$ is the probability of $i^{\rm th}$ individual
receiving an information on the issue $\alpha$, $r$ is a uniform
random number between $0$ and $1$, $o_{i \alpha}$ and $\eta_{i,
  \alpha}$ are the opinion and the tolerance level of the $i^{\rm th}$
individual on the issue $\alpha$ respectively.

At any discrete time step each individual, $i$, may receive a message
on the issue $\alpha$ if he/she is supporting his/her idea less than a
threshold value $ \eta_{i, \alpha}$. The individual adopts the
incoming message, $M_{i \alpha}$ with a probability of $P_{i
  \alpha}$. Hence the opinion value is replaced with a new value which
is modified by the message content. Messages can be sent by either one
of the two parties or by any existing friend. If the $i^{\rm th}$
individual is a supporter of one of the parties but not a committed
follower of its political stands on some of the issues the message may
convert the individual as a new supporter of the opposite party.

 The society can have different
sensitivities on different issues which are represented as
$s_{\alpha}$ in the numerical simulation. To deal with this scenario
three opinion weights can be assigned, $s_{\alpha}\;\; \alpha=1,2,3$, to
each three opinion components (Eq. \ref{resultedopinion}).

\begin{equation}
\label{resultedopinion}
RO_i(t) =   s_1 o_{i 1}(t)+s_2 o_{i 2}(t)+s_3 o_{i 3}(t)
\end{equation}
where $RO_i(t)$ is the resulted opinion of the $i^{\rm th}$
individual, $s_{\alpha}$ and $o_{i\alpha}(t)$ are the
weights,$\alpha=1,2,3$, and the opinions on different issues
respectively.

A binary decision of the individual proceed continuous opinion components
which is calculated by using the sign of the Eq. \ref{resultedopinion}
\ref{decision}.
\begin{equation}
D_i = \left \{ \begin{array}{rl} 1 & if \;\;RO_i > 0 \\  -1 & if \;\; RO_i < 0 \end{array} \right.
\label{decision}
\end{equation}
If the overall opinion of the individual is positive we say
an individual is the supporter of the first view, otherwise the second view.

 Both parties use regular media and social media communications to
 locate weakly committed individuals and try to win them over by
 sending messages.  The system evolves a one-time step in discrete
 time as follows;


\begin{enumerate}
\item Choose randomly an individual,$i$, from the set $\{1,2,\dots,N\}$
\item Choose randomly a  neighbor,$j$, from the set $\{1,2,\dots,N\}$
\item Choose randomly an issue,$\alpha$, to discuss from the set $\alpha=1,2,3$
\item Check the opinion component difference between individual $i$
  and individual $j$ on issues $\alpha$
\item If $diff=o_{i\alpha} -o_{j\alpha} $ is less than tolerance
  threshold,$\Delta$,  exchange opinion with the rule;
\begin{eqnarray}
  o_{i\alpha}(t) &=& \omega o_{i \alpha}(t-1) + (  1- \omega ) o_{j \alpha}(t-1)\nonumber \\
  o_{j\alpha}(t) &=& \omega o_{j\alpha}(t-1) + ( 1- \omega ) o_{i \alpha}(t-1) \nonumber
\end{eqnarray}
\item If $o_{i\beta} < \eta$, where $\beta$ is the issue on which
  external observer send messages,
\item Individual $i$ receive a targeted message $M_{i \beta}$
\item Chose a random number,  $r\in(0,1)$
\item If $P_{\beta}>r$ where $P_{\beta}$ is the probability to adopt
  the message, individual $i$ accepts the message and update opinion
  $o_{i \beta} = o_{i \beta} + M_{i \beta}$
\item Repeat starting from the first step and continue $N$ times.
\end{enumerate}

The above steps describe $1$ discrete time step. The system is
followed until the final date of the campaign.

$\;$

In the next section simulation results, obtained by applying the
the proposed model is introduced with figures.

\section{Results and Discussion}

The proposed model, described in section \ref{Model} contains two
different but complementary interactions among the members of an
artificial society which consists of $N=40000$ fully connected
individuals. Simulations are carried on discrete time steps. A time
step is defined as the number of interactions, $\mathbb{O}(N)$ which
is sufficient for each individual to interact with at least with one
neighbor and one outside news source. At each time slice the averages
are taken over the opinion configurations. In the simulations, 500
different initial opinion configurations are created.  The time span of
the election campaign is chosen as $200$ time steps.

The society, initially, consists of equally divided group of
individuals. Each opinion component has a Gaussian distribution with
mean value $\pm 1$, indicates two opposing views, and variance
$\sigma^2=0.5$. With these choices, the Gaussian opinion distributions
overlap at the origin. As the variance becomes closer to $1$ the
overlapping opinions increases. The individuals who constitutes the
overlap region ( uncommitted supporters of opposing ideas) are the
targeted individuals by the external influences to persuade to their
view.

$\;$

The interaction parameters are grouped into two. The first group is
related to peer-to-peer interactions while the second one is external
influences.
\begin{enumerate}
\item {Peer-to-Peer Interaction Parameters}
Two parameters, the tolerance limit,
$\Delta_{i\alpha}$ and opinion exchange parameter $\omega$ controls the peer-to-peer interactions. $\Delta_{i\alpha}$ is taken as a constant for all
members of the society and all issues, $\Delta_{i\alpha}=\Delta$.  The choice of tolerance parameter $\Delta = 1.216$ allow the individuals to interact with a wide range of opinion holders only excludes extremists. The opinion exchange parameter is taken as, $\Omega = 0.8$ which controls the speed of the opinion formation process. 

\item{The influence of the external sources} \\
  
  The sign of the resultant opinion (Eq.(\ref{resultedopinion})) is the indicator of the vote where the relative weights of the issues are taken
  equal fir the simplicity of the discussions. The conviction parameter,
  $\eta$, is considered as a small value in the simulation studies the
  value is used as $\eta=0.3$. The news acceptance probability,
  $P_{\beta}$, take different values according to the alignment of the
  opinion of the individual and the incoming news item.  The message
  size, $M_{i \beta}$, is also an other parameter which changes at
  each interaction. It is taken as random value, $0 \le M_{i \beta} < 0.5$.
\end{enumerate}

Four different
situations are considered: (a) averaged opinion with only
peer-to-peer interactions ($P_{\beta}=0 ;\;{\rm for}\;\; \beta = 0,1
$), (b) one of the opinion supporters spread information by using
mainstream and social media while the other opinion spread only by
peer-to-peer interactions, ($P_{0}=0.5 ;\; {\rm and}\;\;P_{1}=0.0 $ )
(c) both opinion followers use the same means of external influences,
($P_{\beta}=0.5 ;\;{\rm for}\;\; \beta = 0,1 $) (d) both parties use
influential  sources together with peer-to-peer interactions, but one put more convincing arguments forward,
$P_{0}=1.0 ;\; {\rm and}\;\;P_{1}=0.5 $.

\subsection{Individual interact with others through peer-to-peer interactions no external sources affects the dynamics.}

\begin{figure}
  \begin{center}
    \includegraphics[scale=0.3]{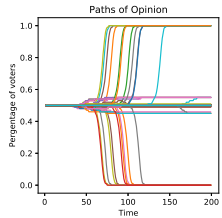}
  \end{center}
  \caption{Paths of opinion change starting from statistically independent initial configurations.}
\label{Local}
\end{figure}

Figure \ref{Local} shows
simulation results starting from different initial opinion configurations
which may be interpreted as the opinion changes in a region during an
election campaign (Paths of opinion). As it can be observed from the
figure, (Figure \ref{Local}) each initial configuration converges a
different final state. This situation resemble election results in different election zones. In different election zones, majority support may
be on different parties but the overall votes are the decisive factor
for the result of the election.

\begin{figure}
  \begin{center}
    \includegraphics[scale=0.3]{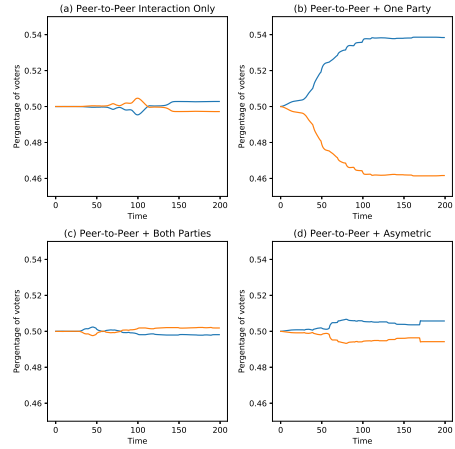}
  \end{center}
  \caption{Time evolution of the configuration averaged opinion distributions.}
\label{Global}
\end{figure}

In fact non of these individual paths has much meaning, the
result of the campaign is the average of all these paths. Figure
\ref{Global} (a) shows that if there is no external influences in
the average both parties share the population almost equally.

\begin{figure}
  \begin{center}
    \includegraphics[scale=0.3]{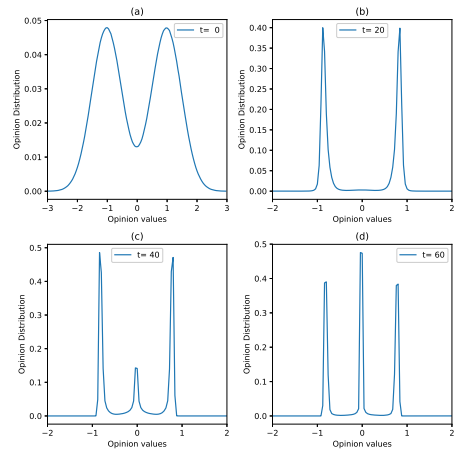}
  \end{center}
  \caption{The snapshots  of the  averaged  opinion distribution without external influences.}
\label{BCOpinionDistribution}
\end{figure}

The dynamics of opinion formation without external influences can be
better understood by observing the changes of the opinion
distributions. Figure \ref{BCOpinionDistribution}, show the averaged
opinion distributions for four instances of the election campaign.

At the initial stages of the campaign, $t=0$ (Figure \ref{BCOpinionDistribution} (a)), the society is
assumed to be equally divided on two opposing opinions. As soon as the
campaign starts, bounded confidence dynamics unite individuals around
the opposing opinions which sharpens the Gaussian opinion
distributions. This situation is not stationary, a third peak start to
appear around the origin, $t=40$ (Figure \ref{BCOpinionDistribution} (c)). As the time passes, the supporters
of both parties, apart from some extremists, converge towards a
moderate opinion ($t=60$, (Figure \ref{BCOpinionDistribution} (d)))  and the distribution remains the same (Figure \ref{FinalDistributions} (a)).

\subsection{\label{OneSided}One of the parties send messages to convert less convinced individuals.}

In any election campaign, ideally, both parties use the
same means to convince individuals. Never the less, it
is not always possible to maintain the same level of publicity or use
of media for both parties.  The uncommitted ($\left|{O_{i,\beta}}\right| < \eta$) electors remain under one-sided news bombardment which may change
opinion of some of the individuals who aim to vote for one party
without a deep conviction. The average daily progress of the opinion
formation results are presented by figure \ref{Global} (b).

\begin{figure}
  \begin{center}
    \includegraphics[scale=0.3]{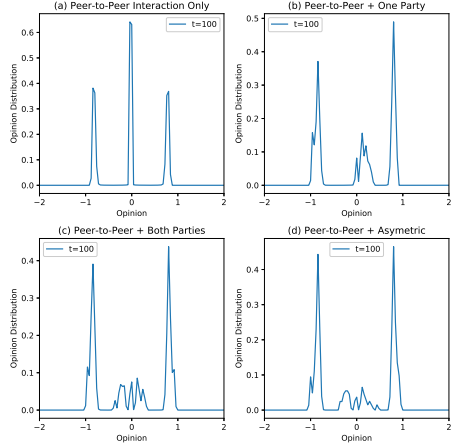}
  \end{center}
  \caption{Configuration averaged opinion distributions .}
\label{FinalDistributions}
\end{figure}

If the outside sources can convince even a small group of uncommitted
individuals it may be sufficient to win the election. The time
dependent variations of the opinion distributions also give clear
picture of the process of opinion changes.  Figure
\ref{FinalDistributions} (b) show that at $t=100$, peer-to-peer
interactions sharpen the Gaussian while the external influences
change the heights of the Gaussian's. At the final stages of the
election campaign the middle peak lean towards the opinion who use
external sources to convince moderate individuals.

\subsection{Both of the parties send messages to convert less convinced individuals.}

Figures \ref{Global} (c) and  \ref{FinalDistribution} (c) show that if both parties are using external news sources and the social media to convince the less committed individuals, the picture is quite similar to the one seen  for only peer-to-peer interaction case. The moderate individuals fluctuate between two opposing opinions, hence, the final election result is unpredictable, figure \ref{Global} (c)

\subsection{Both parties send messages but one sends more convincing (fabricated)  messages. }

The final consideration is that both parties use media and social
media. One of the parties increase activities, send targeted, fake or
fabricated messages, on the social media during the election
period. This resemble the situation during 2016 US Presidential
elections~\cite{Bovet:2018}. The situation is not as sewer as the one
party usage of the social media (Discussed in subsection
\ref{OneSided}) but even such an effort difference can be sufficient
for winning the election. Figure \ref{LocalVsGlobal} show the

Fig.\ref{BCSurveyResultedOpinion} that the percentage of the decision
of two parties under the effect of propaganda by both political
side. This figure shows an alternating behavior and when it comes to
election day the election result can be rather unpredictable.

\section{Conclusions}

Recently the internet is the primary source of acquisition of
knowledge for the societies. This makes the internet a very powerful
and unique. An exciting information, whether it is fake, fabricated or
destructive propagate very fast among the members of the
societies. Media can put forward some ideas or hide some information,
by censoring, to make followers/members gain an advantage.  In social
interactions the spread of gossip and fabricated information has a
very long history and it is not only limited with the online
media~\cite{Soll:2016}.  Never the less the involvements of various
data companies on the 2016 US presidential elections are publicly
known and opened a debate on the violation of civil liberties ~\cite{Bovet:2019}. Such a
data-driven research needs of using an interdisciplinary
approach. Using data science techniques to understand voter behavior
on the segment of their ideas allow politicians to use digital-
marketing strategies to reach individuals. Hence usage of powerful
data analyzing techniques is becoming increasingly harmful to the
civil liberties. As the 2020 US presidential election is approaching
the studies on the effects of the fake and fabricated news and
personalized, targeted messages gain upmost importance.

The present work aims to introduce a simple agent-based model to
simulate the effects of using gathered information to send targeted
messages during an election process. Recent studies show that societies
are almost evenly divided on the main political issues. Hence such a
targeted external information bombardment may be very effective to
change the opinions. Since all parties may use the same techniques, a
small percentage of, $(1\%-2\%)$, opinion fluctuations may be decisive
on the result of the elections. The above assumptions seem reasonably
realistic considering voting processes such as Scottish referendum
$(55.3\% - 44.7\%)$, 2016 US Election $(46.1\% - 48.2\%)$ and Brexit
referendum $(51.9\% - 48.1\%)$. It is observed that if both parties
compete equally the election results are unpredictable. If one of the
parties use the technological power and social media more than the
opponents, can easily gain the required small percentage of the
undecided voters. It is evident that in the next decade the use of artificial
intelligence techniques to extract information from individuals social
media history will be used more frequently unless some global
legislative regulation on the use of private information.

\end{document}